\documentclass[a4paper,fleqn,usenatbib]{article}

\usepackage{ae,aecompl}

\usepackage[english]{babel}
\usepackage[T1]{fontenc}
\usepackage{newtxtext,newtxmath}

\usepackage[a4paper,top=3cm,bottom=2cm,left=3cm,right=3cm,marginparwidth=1.75cm]{geometry}

\usepackage{amsmath}
\usepackage{graphicx}
\usepackage[colorinlistoftodos]{todonotes}
\usepackage[colorlinks=true, allcolors=blue]{hyperref}
\usepackage{amssymb}	
\usepackage{multirow}
\usepackage[left]{lineno}
\usepackage{booktabs}
\usepackage{natbib}

\title{\huge{Search for high-energy neutrinos in coincidence with Fast Radio Bursts with the ANTARES neutrino telescope}}
\author{The ANTARES Collaboration \and
        A.~Albert$^{1}$, 
		M.~Andr\'e$^{2}$, 
		M.~Anghinolfi$^{3}$, 
		G.~Anton$^{4}$, 
		M.~Ardid$^{5}$, 
		J.-J.~Aubert$^{6}$, 
		J.~Aublin$^{7}$, \and 
		T.~Avgitas$^{7}$, 
		B.~Baret$^{7}$, 
		J.~Barrios-Mart\'{\i}$^{8}$
% ...
\\
}

\date{\normalsize{Received \today; accepted ...}}

\begin{document}
\maketitle

$^{1}$\scriptsize{Universit\'e de Strasbourg, CNRS,  IPHC UMR 7178, F-67000 Strasbourg, France}\\
$^{2}$\scriptsize{Technical University of Catalonia, Laboratory of Applied Bioacoustics, Rambla Exposici\'o, 08800 Vilanova i la Geltr\'u, Barcelona, Spain}\\
$^{3}$\scriptsize{INFN - Sezione di Genova, Via Dodecaneso 33, 16146 Genova, Italy}\\
$^{4}$\scriptsize{Friedrich-Alexander-Universit\"at Erlangen-N\"urnberg, Erlangen Centre for Astroparticle Physics, Erwin-Rommel-Str. 1, 91058 Erlangen, Germany}\\
$^{5}$\scriptsize{Institut d'Investigaci\'o per a la Gesti\'o Integrada de les Zones Costaneres (IGIC) - Universitat Polit\`ecnica de Val\`encia. C/  Paranimf 1, 46730 Gandia, Spain}\\
$^{6}$\scriptsize{Aix Marseille Univ, CNRS/IN2P3, CPPM, Marseille, France}\\
$^{7}$\scriptsize{APC, Univ Paris Diderot, CNRS/IN2P3, CEA/Irfu, Obs de Paris, Sorbonne Paris Cit\'e, France}\\
$^{8}$\scriptsize{IFIC - Instituto de F\'isica Corpuscular (CSIC - Universitat de Val\`encia) c/ Catedr\'atico Jos\'e Beltr\'an, 2 E-46980 Paterna, Valencia, Spain}\\
$^{9}$\scriptsize{LAM - Laboratoire d'Astrophysique de Marseille, P\^ole de l'\'Etoile Site de Ch\^ateau-Gombert, rue Fr\'ed\'eric Joliot-Curie 38,  13388 Marseille Cedex 13, France}\\
$^{10}$\scriptsize{National Center for Energy Sciences and Nuclear Techniques, B.P.1382, R. P.10001 Rabat, Morocco}\\
$^{11}$\scriptsize{INFN - Laboratori Nazionali del Sud (LNS), Via S. Sofia 62, 95123 Catania, Italy}\\
$^{12}$\scriptsize{Nikhef, Science Park,  Amsterdam, The Netherlands}\\
$^{13}$\scriptsize{Huygens-Kamerlingh Onnes Laboratorium, Universiteit Leiden, The Netherlands}
$^{14}$\scriptsize{University Mohammed V in Rabat, Faculty of Sciences, 4 av. Ibn Battouta, B.P. 1014, R.P. 10000
	Rabat, Morocco}\\
$^{15}$\scriptsize{Institute of Space Science, RO-077125 Bucharest, M\u{a}gurele, Romania}\\
$^{16}$\scriptsize{Universiteit van Amsterdam, Instituut voor Hoge-Energie Fysica, Science Park 105, 1098 XG Amsterdam, The Netherlands}\\
$^{17}$\scriptsize{INFN - Sezione di Roma, P.le Aldo Moro 2, 00185 Roma, Italy}\\
$^{18}$\scriptsize{Dipartimento di Fisica dell'Universit\`a La Sapienza, P.le Aldo Moro 2, 00185 Roma, Italy}\\
$^{19}$\scriptsize{Gran Sasso Science Institute, Viale Francesco Crispi 7, 00167 L'Aquila, Italy}\\
$^{20}$\scriptsize{LPHEA, Faculty of Science - Semlali, Cadi Ayyad University, P.O.B. 2390, Marrakech, Morocco.}\\
$^{21}$\scriptsize{INFN - Sezione di Bologna, Viale Berti-Pichat 6/2, 40127 Bologna, Italy}\\
$^{22}$\scriptsize{INFN - Sezione di Bari, Via E. Orabona 4, 70126 Bari, Italy}\\
$^{23}$\scriptsize{Department of Computer Architecture and Technology/CITIC, University of Granada, 18071 Granada, Spain}\\
$^{24}$\scriptsize{G\'eoazur, UCA, CNRS, IRD, Observatoire de la C\^ote d'Azur, Sophia Antipolis, France}\\
$^{25}$\scriptsize{Dipartimento di Fisica dell'Universit\`a, Via Dodecaneso 33, 16146 Genova, Italy}\\
$^{26}$\scriptsize{Universit\'e Paris-Sud, 91405 Orsay Cedex, France}\\
$^{27}$\scriptsize{University Mohammed I, Laboratory of Physics of Matter and Radiations, B.P.717, Oujda 6000, Morocco}\\
$^{28}$\scriptsize{Institut f\"ur Theoretische Physik und Astrophysik, Universit\"at W\"urzburg, Emil-Fischer Str. 31, 97074 W\"urzburg, Germany}\\
$^{29}$\scriptsize{Dipartimento di Fisica e Astronomia dell'Universit\`a, Viale Berti Pichat 6/2, 40127 Bologna, Italy}\\
$^{30}$\scriptsize{Laboratoire de Physique Corpusculaire, Clermont Universit\'e, Universit\'e Blaise Pascal, CNRS/IN2P3, BP 10448, F-63000 Clermont-Ferrand, France}\\
$^{31}$\scriptsize{LIS, UMR Universit\'e de Toulon, Aix Marseille Universit\'e, CNRS, 83041 Toulon, France}\\
$^{32}$\scriptsize{Royal Netherlands Institute for Sea Research (NIOZ) and Utrecht University, Landsdiep 4, 1797 SZ 't Horntje (Texel), the Netherlands}\\
$^{33}$\scriptsize{Institut Universitaire de France, 75005 Paris, France}\\
$^{34}$\scriptsize{Dr. Remeis-Sternwarte and ECAP, Friedrich-Alexander-Universit\"at Erlangen-N\"urnberg,  Sternwartstr. 7, 96049 Bamberg, Germany}\\
$^{35}$\scriptsize{Moscow State University, Skobeltsyn Institute of Nuclear Physics, Leninskie gory, 119991 Moscow, Russia}\\
$^{36}$\scriptsize{Mediterranean Institute of Oceanography (MIO), Aix-Marseille University, 13288, Marseille, Cedex 9, France; Universit\'e du Sud Toulon-Var,  CNRS-INSU/IRD UM 110, 83957, La Garde Cedex, France}\\
$^{37}$\scriptsize{INFN - Sezione di Catania, Via S. Sofia 64, 95123 Catania, Italy}\\
$^{38}$\scriptsize{Direction des Sciences de la Mati\`ere - Institut de recherche sur les lois fondamentales de l'Univers - Service de Physique des Particules, CEA Saclay, 91191 Gif-sur-Yvette Cedex, France}\\
$^{39}$\scriptsize{INFN - Sezione di Pisa, Largo B. Pontecorvo 3, 56127 Pisa, Italy}\\
$^{40}$\scriptsize{Dipartimento di Fisica dell'Universit\`a, Largo B. Pontecorvo 3, 56127 Pisa, Italy}\\
$^{41}$\scriptsize{INFN - Sezione di Napoli, Via Cintia 80126 Napoli, Italy}\\
$^{42}$\scriptsize{Dipartimento di Fisica dell'Universit\`a Federico II di Napoli, Via Cintia 80126, Napoli, Italy}\\
$^{43}$\scriptsize{Dpto. de F\'\i{}sica Te\'orica y del Cosmos \& C.A.F.P.E., University of Granada, 18071 Granada, Spain}\\
$^{44}$\scriptsize{GRPHE - Universit\'e de Haute Alsace - Institut universitaire de technologie de Colmar, 34 rue du Grillenbreit BP 50568 - 68008 Colmar, France}
\\

\normalsize
\begin{abstract}
In the past decade, a new class of bright transient radio sources with millisecond duration has been discovered. The origin of these so-called Fast Radio Bursts (FRBs) is still a great mystery despite the growing observational efforts made by various multi-wavelength and multi-messenger facilities. So far, many models have been proposed to explain FRBs but neither the progenitors nor the radiative and the particle acceleration processes at work have been clearly identified. In this paper, the question whether some hadronic processes may occur in the vicinity of the FRB source is assessed. If so, FRBs may contribute to the high energy cosmic-ray and neutrino fluxes. A search for these hadronic signatures has been done using the ANTARES neutrino telescope. The analysis consists in looking for high-energy neutrinos, in the TeV-PeV regime, spatially and temporally coincident with the detected FRBs. Most of the FRBs discovered in the period 2013-2017 were in the field of view of the ANTARES detector, which is sensitive mostly to events originating from the Southern hemisphere. From this period, 12 FRBs have been selected and no coincident neutrino candidate was observed.
Upper limits on the per burst neutrino fluence have been derived using a power law spectrum, $\rm{\frac{dN}{dE_\nu}\propto E_\nu^{-\gamma}}$, for the incoming neutrino flux, assuming spectral indexes $\gamma$ = 1.0, 2.0, 2.5. Finally, the neutrino energy has been constrained by computing the total energy radiated in neutrinos assuming different distances for the FRBs. Constraints on the neutrino fluence and on the energy released are derived from the associated null results.
\end{abstract}


\section{Introduction}
\label{sec_intro}

Discovered in the last decade \citep{Lorimer07}, Fast Radio Bursts (FRBs) are characterised by a short duration (t $\sim$ a few ms) of intense radio emission ($\rm{>1~Jy\cdot ms}$) measured so far in the 800 MHz and 1.4 GHz bands by various radio telescopes. The astrophysical origin of FRBs is largely unknown. However, their high dispersion measures (DM), due to the scattering of the radio wave propagating through an ionised column of matter, suggest an extragalactic/cosmological origin \citep{Lorimer05,Thornton13}.\\ 
From the 34 FRBs already detected and reported in the FRB catalog\footnote{see the FRB catalog : \url{http://frbcat.org/}} \citep{Petroff16}, the observed DM can be used to derive upper limits on their cosmological redshift, $z_{DM} \in [0.12 ~; 2.3]$. This translates into an upper limit on the isotropic radio energy release of $E_{rad} \in [10^{38} ~; 10^{41}]$ erg. Up to now, the FRB progenitors are thought to originate from a large variety of astrophysical sources \citep{Keane16} usually split into two classes: the repeating ones and the single cataclysmic events. Indeed, such a large amount of energy, released in a millisecond timescale, may favor an FRB origin from violent cataclysmic events. Those are powered by compact objects where the progenitor does not survive afterwards (single burst model). Several models have been proposed such as neutron star mergers \citep{Totani13, Wang16} possibly associated to short Gamma-Ray Bursts (GRBs) \citep{Zhang14, Palaniswamy14,Murase17} or supramassive neutron star collapses \citep{Falcke14,Ravi14, Li14}. Non-destructive flaring models including giant pulses from young and rapidly rotating neutron stars \citep{Pen15,Cordes16}, magnetar giant flares \citep{Popov13,Lyubarsky14}, hyperflares from soft gamma-repeaters \citep{Popov10}, a young neutron star embeded in a wind bubble \citep{Murase16} or maybe from the interior of young supernovae \citep{Bietenholz17} are however good astrophysical candidates to explain both types of repeating and non repeating FRBs. More exotic models have also been proposed such as radio burst radiation of superconducting cosmic strings \citep{Cao18,Ye17}. \\
Recently, the discovery of the repeating behavior of FRB 121102 \citep{Spitler16, Scholz16} has brought new insights into the nature of the FRB progenitors. In addition, radio interferometric observations of FRB 121102 \citep{Marcote17,Chatterjee17} made possible, for the first time, to unambiguously determine the redshift of the FRB source at $z\sim0.19$ \citep{Tendulkar17}, confirming the tremendous amount of radio energy that can be released during an FRB event.\\
Radio observation campaigns have been done to search for other FRB "repeaters" among the known FRB population but without any success \citep{Lorimer07,Ravi15,Petroff15,Petroff17}. However, the instrumental sensitivities and the short observation time of a few hours may account for this null result. Therefore, the question whether FRB 121102 belongs or not to a special class of FRB is still under debate.\\ 
The spatial distribution and the all-sky rate of FRBs, $\rm{R_{FRB}}$, can provide additional constraints on the nature of the FRB progenitors when it is compared to those of known astrophysical sources. The all-sky rate  $\rm{R_{FRB} \sim 10^3~day^{-1}}$ has been estimated for radio pulses with $\rm{F > 1~Jy \cdot ms}$ \citep{Champion16}. This high event rate would already rule out a short GRB-dominated population of FRBs since $\rm{R_{FRB}/R_{short~GRB}\sim 10^3}$ assuming that $\rm{R_{short~GRB} = \frac{N_{short~GRB}}{N_{all~GRB}}\times R_{GRB}}$ where $\rm{R_{all~GRB} = 1000~yr^{-1}}$ in the entire sky and the detected GRB population is composed of 1/3 of short GRBs according to the CGRO-BATSE observations \citep{Goldstein13}. Alternatively, $\rm{R_{FRB}}$ corresponds to only 10\% of the observed CCSNe rate \citep{Thornton13}. Therefore, the CCSNe reservoir may account for the high event rate of FRBs. For instance, \cite{Falcke14} claimed that only 3\% of the core collapse supernovae (CCSNe) producing supramassive neutron stars are needed to explain the FRB rate.
The various models proposed are difficult to discriminate because of lack of additional information on the broadband FRB spectra. Many multi-wavelength follow-ups have been organised recently \citep{Petroff14,Petroff17,Scholz17,Bhandari17,Hardy17} but no counterpart (optical/x-rays/gamma-rays/VHE gamma-rays) has been identified yet. However, in 2016, \cite{DeLaunay16} reported the detection of a gamma-ray GRB-like counterpart in association with FRB 131104 but with a small significance (3.2 $\sigma$). For FRB 131104, \cite{DeLaunay16} determined that the radio to gamma-ray energy output ratio would be $\rm{E_{rad}/E_{\gamma} >10^{-9}}$ assuming the source is at the redshift inferred by the DM measurement. This may show that a large fraction of the total energy radiated during these radio bursting events may be emitted at high energy while being still undetected or marginally detected.  
If the radio emission is likely produced by coherent emission of leptons \citep[][and references therein]{Katz14}, hadronic processes may be the source of the most energetic photons in the gamma-ray energy domain. In this case, TeV-PeV neutrinos can be produced by photohadronic interactions. These hadronic processes may occur in the energetic outflow released during a cataclysmic FRB event \citep{Falcke14} or in the vicinity of the FRB progenitor through the interaction of the outflow with the surrounding environment \citep{Zhang03,Li14,Murase16,Dey16}.\\
Based on their high rate, $\rm{R_{FRB}}$, and under the assumption that a fraction of FRBs are indeed efficient accelerators of TeV-PeV hadrons, they may contribute significantly to the cosmic diffuse neutrino signal discovered by the IceCube Collaboration \citep{Aartsen13,Aartsen15a,Aartsen15b,Aartsen15c,Aartsen16}.
This diffuse astrophysical neutrino signal is now established with a high significance. The ANTARES neutrino telescope also observes a mild excess over the background of neutrino candidates at high energies \citep{ANTARES17}. Up to now, no population of astrophysical sources clearly emerge from the background to explain this diffuse flux. However, recently, the IceCube Collaboration claimed the evidence of a high-energy neutrino signal from the blazar TXS 0506+056 \citep{IC18a,IC18b} which marks an additional step towards the identification of the nature of the cosmic accelerators in the Universe.
Multi-messenger observations of FRBs are crucial to probe them as cosmic accelerators. So far, neutrino searches from FRBs by the IceCube \citep{Fahey17, IC17} and the ANTARES \citep{ANTARES17_2} Collaborations yielded a null result.\\
In this paper, a search for neutrinos in coincidence with FRBs detected between 2013 and 2017 using the ANTARES neutrino telescope is presented. Located in the Mediterranean Sea, ANTARES is the largest high-energy neutrino telescope in the Northern Hemisphere, operating since 2008 \citep{Ageron11}. By design, the ANTARES detector continuously monitors, with a high duty cycle and good angular resolution, the Southern sky (2$\pi$ steradian at any time), where most of the FRBs have been discovered to date. In section \ref{sec:search}, the FRB sample used in the analysis is described as well as the results of the search for a neutrino counterpart. The constraints on the neutrino fluence and energy emission are given in section \ref{sec:constraints}. Finally, in section \ref{sec:Discussion}, the results are discussed with respect to different expectations from FRB hadronic models and the FRB contribution to the diffuse neutrino flux. The conclusions are drawn in section \ref{Conclusion}.

\section{Search for high-energy neutrinos from FRBs in the ANTARES data}
\label{sec:search}

\subsection{The FRBs in the field of view of ANTARES}

\begin{figure}
	\centering
	\includegraphics[trim= 0 0 0 0,clip=true,width = 0.7\columnwidth]{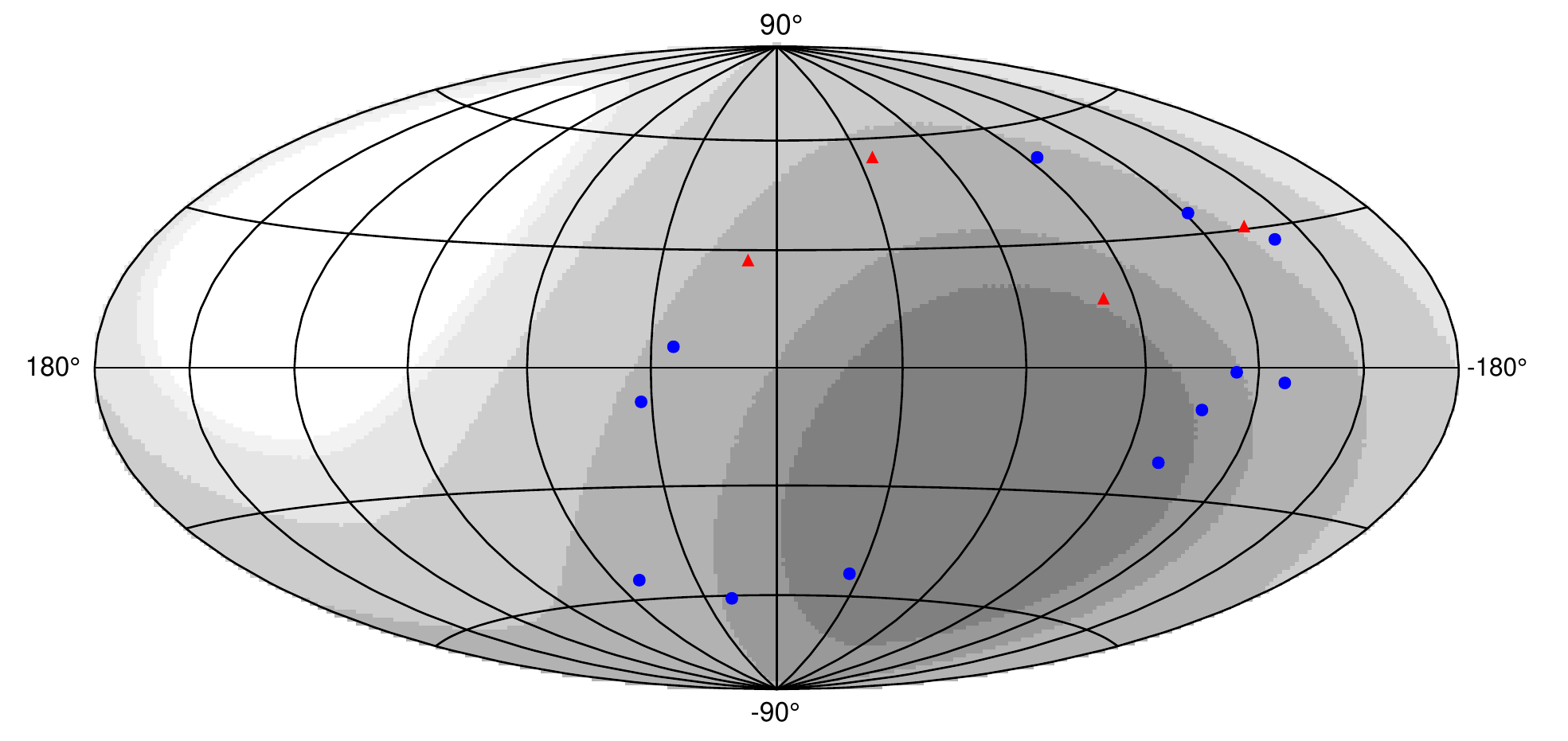}
	\caption{Skymap in Galactic coordinates showing the positions of the 16 FRBs detected in the period 2013-2017. The 12 selected FRBs are shown with the blue dots while the 4 non-selected FRBs are displayed with the red triangles. The region of the sky observable by ANTARES (on average) is also displayed in greyscale from 100\% of visibility for the darkest area to 0\% for the white area when considering upgoing neutrino candidates in the detector. }
	\label{fig:sky_map}
\end{figure}

This analysis focuses on the period from Jan. 2013 to Jan. 2017 during which 16 FRBs\footnote{During the review of this paper, a 17th FRB (FRB 141113) has been reported by \cite{Patel18}. Occuring in 2014, FRB141113 was below the ANTARES horizon up to 2 hours after its trigger time.} were detected by the Parkes telescope, UTMOST and ASKAP. When active, the ANTARES telescope monitors the sky region with declinations $\delta < -48^\circ$ with almost 100\% duty cycle; for $-48^\circ <\delta < +48^\circ$ the duty cycle decreases gradually because of the requirement that the neutrino candidates are upgoing. The first selection criterion is that the FRB position must be within the ANTARES field of view (FoV) within a chosen time window $\Delta T = [T_0-6h ~; T_0+6h]$ where $T_0$ is the FRB trigger time. Three FRBs did not fulfill this first selection criterion and were then removed from the sample used in this analysis. In addition, the quality of the ANTARES data acquired during the whole day around each FRB detection was verified to avoid any anomalous behavior of the detector. One more FRB (FRB 150610) was excluded since the detector was not active due to a power cut that happenned 4 hours before the trigger time. At the end, the final sample is composed of 12 FRBs for which ANTARES data were considered.  In table \ref{tab:tab1}, the main properties of the FRB sample are summarised and a sky map of the FRB positions superimposed with the ANTARES sky visibility is shown in figure \ref{fig:sky_map}.

\begin{table}
	\centering
	\caption{Properties of the 12 FRBs visible by ANTARES in the period 2013-2017 according to the FRB catalogue \citep{Petroff16}. $z_{DM}$ corresponds to the upper limit on the cosmological redshift inferred from the DM measured in excess to the Galactic contribution.}
	\begin{tabular}{cccccc}
		   &   &  &  &  & \\
		\hline
		FRB & $z_{DM}$ & $T_0$ & RA & dec  & radio telescope\\
		&               &    (UTC)  & ($\rm{^o}$) & ($\rm{^o}$) & \\
		\hline
		\hline
		\\
		131104 & 0.59  & 18:04:11.20 & 101.04 & -51.28 & Parkes\\
		140514 & 0.44  & 17:14:11.06 & 338.52 & -12.31 & Parkes\\
		150215 & 0.55  & 20:41:41.71 & 274.36 & -4.90  & Parkes\\
		150418 & 0.49  & 04:29:06.66 & 109.15 & -19.01 & Parkes\\
		150807 & 0.59  & 17:53:55.83 & 340.10 & -55.27 & Parkes\\
		151206 & 1.385 & 06:17:52.78 & 290.36 & -4.13  & Parkes\\
		151230 & 0.76  & 16:15:46.53 & 145.21 & -3.45  & Parkes\\
		160102 & 2.13  & 08:28:39.37 & 339.71 & -30.18 & Parkes\\
		160317 & 0.70  & 09:00:36.53 & 118.45 & -29.61 & UTMOST\\
		160410 & 0.18  & 08:33:39.68 & 130.35 &   6.08 & UTMOST\\
		160608 & 0.37  & 03:53:01.09 & 114.17 & -40.78 & UTMOST\\
		170107 & 0.48  & 20:05:45.14 & 170.79 & -5.02  & ASKAP\\
		\hline
	\end{tabular}
	\label{tab:tab1}
\end{table}

\subsection{Method}
\label{subsec:method}
For each selected FRB, the ANTARES data set is extracted within a time window $\Delta T = [T_0-6h~; T_0+6h]$. This time window is chosen to encompass various delay scenarios between the radio and the neutrino signals while keeping the background noise at a low level. Within $\Delta T$, the event rates are checked to verify the detector stability. No significant time variability of the counting rates was found which ensures the quality of the extracted data.
The search for a significant neutrino flux is then based on the detection of upgoing neutrino-induced muons coincident with the position of the FRB within $\Delta T$.\\
To suppress the atmospheric muon background contamination in the neutrino sample, selection cuts are applied using the quality variables of the track reconstruction algorithm \citep{ANTARES12}: the reconstructed zenith angle, $\theta$, the error estimate on the reconstructed direction, $\beta$, and the quality fit parameter, $\Lambda$. Each selected upgoing (cos$\theta >\rm{0}$) event was required to have a direction error $\beta < \rm{1^o}$.  The final selection criteria is based on the quality fit parameter $\Lambda$. For $\Delta T$ centered on each FRB time, the optimal value, $\Lambda_{3\sigma}$, was chosen in such a way that the presence of one neutrino candidate in the time window would correspond to a positive signal with 3$\sigma$ significance \citep{ANTARES17_3}.
Finally, a search cone of 2$^\circ$ is set around each FRB position. From radio information, the typical localization errors corresponds to radii of 10 arcmin. 

\subsection{Results}

No upgoing events spatially and temporally correlated with the 12 selected FRBs were found. This null result is compatible with the background event rate of ANTARES estimated to be $\rm{\sim 5 \cdot 10^{-8}~event \cdot s^{-1}}$.
Since no neutrino signal is detected in coincidence with any of the selected FRBs, constraints on the fluence of neutrinos that would have been observed by the ANTARES detector are derived.

\section{Constraints on the neutrino fluence and energy emission from Fast Radio Bursts}
\label{sec:constraints}
\subsection{Constraints on the per burst neutrino fluence}
An upper limit on the neutrino fluence is computed on a per burst basis with the following procedure. 
For a given neutrino flux, the number of expected events, ${N_\nu}$, depends on the detector effective area, ${A_{eff}(E_\nu,\delta)}$ (units: cm$^{2}$).
Once the selection parameters $\rm{(\cos\theta, \beta,\Lambda_{3\sigma})}$ have been defined, as explained in \S \ref{subsec:method}, the effective area only depends on the neutrino energy, ${E_\nu}$ and the source declination, $\delta$.
To compute the effective area at any declination, dedicated Monte Carlo simulations reproducing the ANTARES data taking conditions at the FRB trigger time, ${T_0}$, have been produced. \\
For a given time-integrated neutrino flux, $\frac{dN}{dE_\nu}$ (units: GeV$^{-1}$ cm$^{-2}$), the number of expected neutrino events for a source at declination $\delta$ is 

\begin{equation}\label{eq2}
N_\nu(\delta) = \int_{E_\nu} {A_{eff}(E_\nu,\delta)}\cdot \frac{dN}{dE_\nu} \cdot dE_\nu \ .
\end{equation}
%
Usually, a neutrino power-law $dN/dE_\nu \propto E_\nu^{-\gamma}$ is assumed. The neutrino fluence at the detector can thus be defined as

\begin{equation}\label{eq2b}
{E^2_\nu} {\frac{dN}{dE_\nu}}=\phi_0 \cdot \biggl( \frac{E_\nu}{E_0} \biggr)^{-\gamma+2} \quad \rm{(in~GeV\cdot cm^{-2})} \ .
\end{equation}
The normalization factor, $\phi_0$, has the same units as the neutrino fluence and it corresponds to the expected neutrino energy fluence at the reference energy, ${E_0}$ = 100 TeV. 
Due to the strong energy-dependence of the effective area with $E_\nu$, the sensitivity of the detector to a given neutrino flux is strongly dependent on the spectral index $\gamma$.
As the neutrino production mechanisms for FRBs are unknown, three spectral models have been tested in this analysis to conservatively cover a large range of possibilities: 
a hard spectrum with $\gamma$ = 1.0 usually considered in some stages of p$\gamma$ acceleration processes, an intermediate spectrum with $\gamma$ = 2.0 corresponding to the theoretical index for Fermi acceleration processes and a softer spectrum with $\gamma$ = 2.5. The latter almost corresponds to the best fit value of the isotropic astrophysical neutrino signal measured by IceCube \citep{Aartsen15b}.

By inverting equation \ref{eq2}, a null neutrino detection can be translated to an upper limit on the normalization factor $\phi_0$ of the energy spectrum for the given values of the spectral index $\gamma$.
Assuming Poisson statistics, the 90\% confidence level (C.L.) upper limit, $\phi_0^{90\%}$, is defined by setting $N_\nu(\delta)=2.3$.
No neutrino event was observed in a temporal coincidence within $T_0\pm 6$h for any of the considered FRBs. 
The values of $\phi_0^{90\%}$ for each FRB and for the three assumed spectral indexes are reported in table \ref{tab:fluence_res}.

Starting from the upper limit on the normalization factor, the corresponding 90\% C.L. upper limits on the fluence for each FRB has also been computed as:

\begin{equation}\label{eq2ter}
F_\nu^{90\%} = \int_{E_{min}}^{E_{max}} E_\nu \cdot \frac{dN}{dE_\nu}\cdot dE_\nu 
= \phi_0^{90\%}\cdot E_0^{\gamma-2} \int_{E_{min}}^{E_{max}} E_\nu^{1-\gamma}\cdot dE_\nu \ .
\end{equation}
The integration is over the range from $E_{min}$ to $E_{max}$, which corresponds to the energies that define the $\rm{5-95\%}$ range of the energy distribution of simulated events passing all the analysis cuts for the corresponding spectrum.

\begin{figure}
	\centering
	\includegraphics[trim= 0 150 0 180,clip=true,width =0.8\columnwidth]{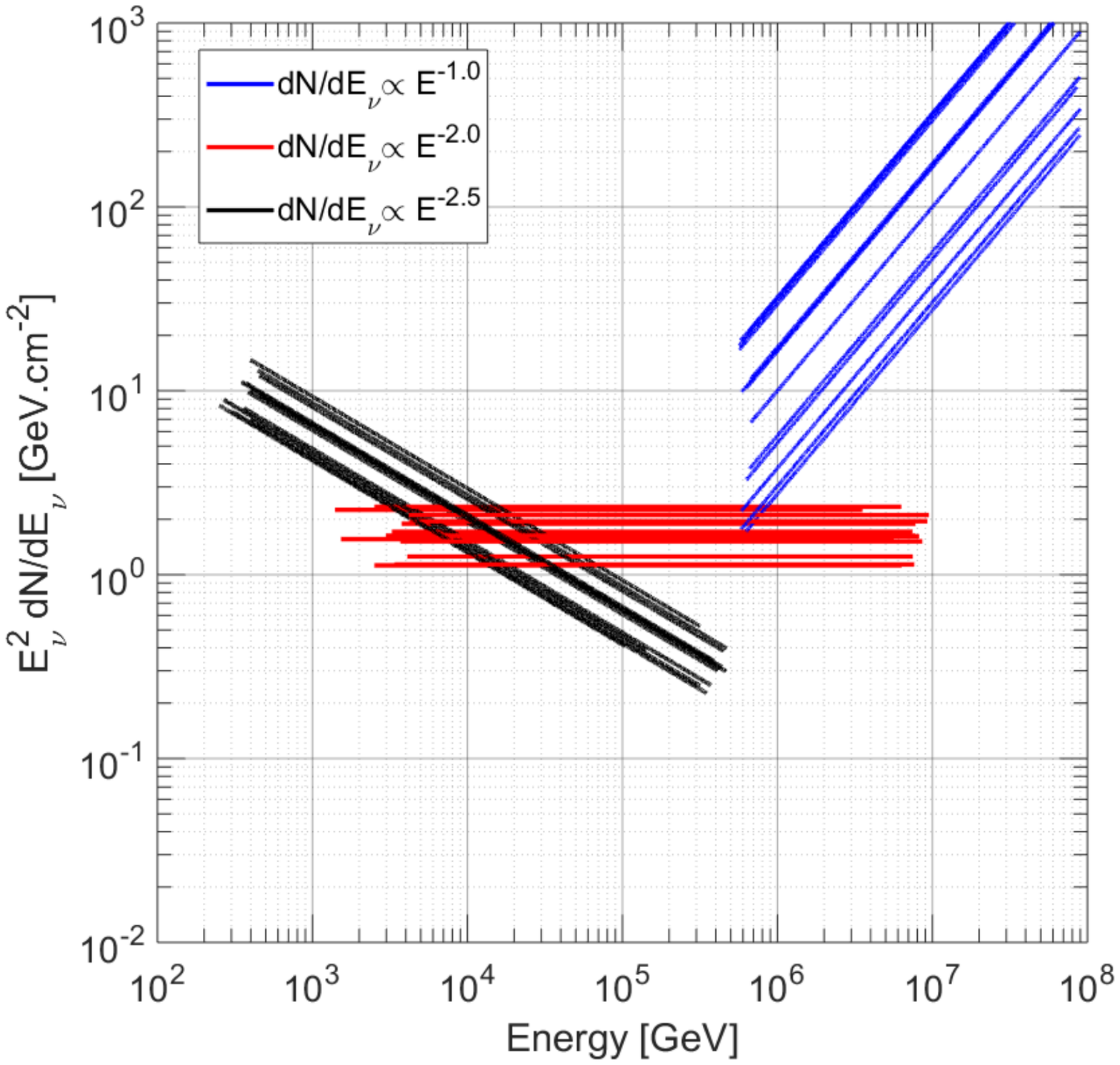}
	\caption{The 90\% confidence level ANTARES upper limits on the neutrino fluence for the power law spectral models with $\gamma$ = 1.0 (blue), 2.0 (red), 2.5 (black), for each FRB. The limits are computed in the energy range [$E_{min}~;E_{max}$] where 5-95\% of the neutrino signal is expected.}
	\label{fig:limit_fluence}
\end{figure}

The upper limits on the neutrino fluence, $F_\nu^{90\%}$, for each individual FRB and for the three test spectral indexes are reported in table \ref{tab:fluence_res} and shown in figure \ref{fig:limit_fluence}. As shown later in figure \ref{fig:Aeff_all}, compared to a time-integrated neutrino point source analysis \citep[e.g.][]{Albert17c}, searching for neutrinos from short transient events permits to improve the upper limit derived on the neutrino fluence by a factor of $\sim$ 30\%.

\begin{table*}
	
	\caption{The 90$\%$ C.L upper limits on the spectral fluence, $\phi_0^{90\%}$, evaluated at 100 TeV, and the fluence, $F_\nu^{90\%}$, for the three spectral models considered in this analysis ($\gamma$ = 1.0, 2.0, 2.5). $\phi_0^{90\%}$ and $F_\nu^{90\%}$ are expressed in $\rm{GeV \cdot cm^{-2}}$. The [$5\%~;95\%$] energy boundaries, $E_{min}$ and $E_{max}$, used to compute the energy integrated fluence are also shown in the units of $\rm{log_{10}(E_\nu/GeV)}$.}
	\begin{tabular}{c||cccc||cccc||cccc}
		\multicolumn{13}{c}{}  \\
		\hline
		\multirow{ 4}{*}{FRB}	& \multicolumn{4}{c||}{$\gamma =1.0$} & \multicolumn{4}{c||}{$\gamma =2.0$} & \multicolumn{4}{c}{$\gamma =2.5$}  \\ 
		&  &  &  &  &  &  &  &  &  &  &  &   \\
	    & $\phi_0^{90\%}$ & $F_\nu^{90\%}$ & $E_{min}$ & $E_{max}$ & $\phi_0^{90\%}$ & $F_\nu^{90\%}$ & $E_{min}$ & $E_{max}$ & $\phi_0^{90\%}$ & $F_\nu^{90\%}$ & $E_{min}$ & $E_{max}$ \\ 
		&  &  &  &  &  &  &  &  &  &  &  &   \\
		\hline
		\hline
		&  &  &  &  &  &  &  &  &  &  &  &   \\
		131104 & 1.6 & $1.3\cdot10^{3}$ & 5.8 & 7.9 & 1.1 & 8.80 & 3.4 & 6.8 & 0.4 & 13.5 & 2.6 & 5.5  \\ 
		140514 & 3.2 & $2.8\cdot10^{3}$ & 5.8 & 7.9 & 1.9 & 14.4 & 3.6 & 6.9 & 0.6 & 19.1 & 2.6 & 5.6  \\   
		150215 & 2.9 & $2.6\cdot10^{3}$ & 5.8 & 7.9 & 2.3 & 17.7 & 3.1 & 6.5 & 0.4 & 15.8 & 2.4 & 5.0  \\   
		150418 & 1.7 & $1.5\cdot10^{3}$ & 5.8 & 8.0 & 1.7 & 13.2 & 3.5 & 6.9 & 0.5 & 15.6 & 2.6 & 5.6  \\     
		150807 & 0.4 & $3.4\cdot10^{2}$ & 5.8 & 8.0 & 1.6 & 12.1 & 3.6 & 6.9 & 0.9 & 25.1 & 2.6 & 5.7  \\   
		151206 & 0.3 & $2.5\cdot10^{2}$ & 5.8 & 8.0 & 1.3 &  9.4 & 3.6 & 6.9 & 0.7 & 21.7 & 2.5 & 5.6  \\  
		151230 & 1.0 & $8.9\cdot10^{2}$ & 5.8 & 8.0 & 1.6 & 12.8 & 3.2 & 6.8 & 0.5 & 17.1 & 2.4 & 5.2  \\   
		160102 & 0.6 & $5.1\cdot10^{2}$ & 5.8 & 8.0 & 2.0 & 15.0 & 3.6 & 7.0 & 0.8 & 23.4 & 2.7 & 5.7  \\   
		160317 & 3.1 & $2.7\cdot10^{3}$ & 5.8 & 7.9 & 1.6 & 12.8 & 3.5 & 6.9 & 0.4 & 14.9 & 2.5 & 5.5  \\   
		160410 & 0.5 & $4.4\cdot10^{2}$ & 5.8 & 7.9 & 1.5 & 11.8 & 3.6 & 6.9 & 0.6 & 19.6 & 2.6 & 5.6  \\   
		160608 & 1.7 & $1.5\cdot10^{3}$ & 5.8 & 7.9 & 2.1 & 16.3 & 3.6 & 7.0 & 0.6 & 18.5 & 2.7 & 5.7  \\   
		170107 & 0.3 & $2.7\cdot10^{2}$ & 5.8 & 7.9 & 1.1 & 8.8 & 3.5 & 6.9 & 0.7 & 21.3 & 2.6 & 5.6  \\   
		\hline
	\end{tabular}
	\label{tab:fluence_res}
\end{table*}

\subsection{Constraints on the TeV-PeV neutrino energy released by FRBs}

\begin{figure}
	\centering
	\includegraphics[trim= 0 0 0 30,clip=true, width =0.8 \columnwidth]{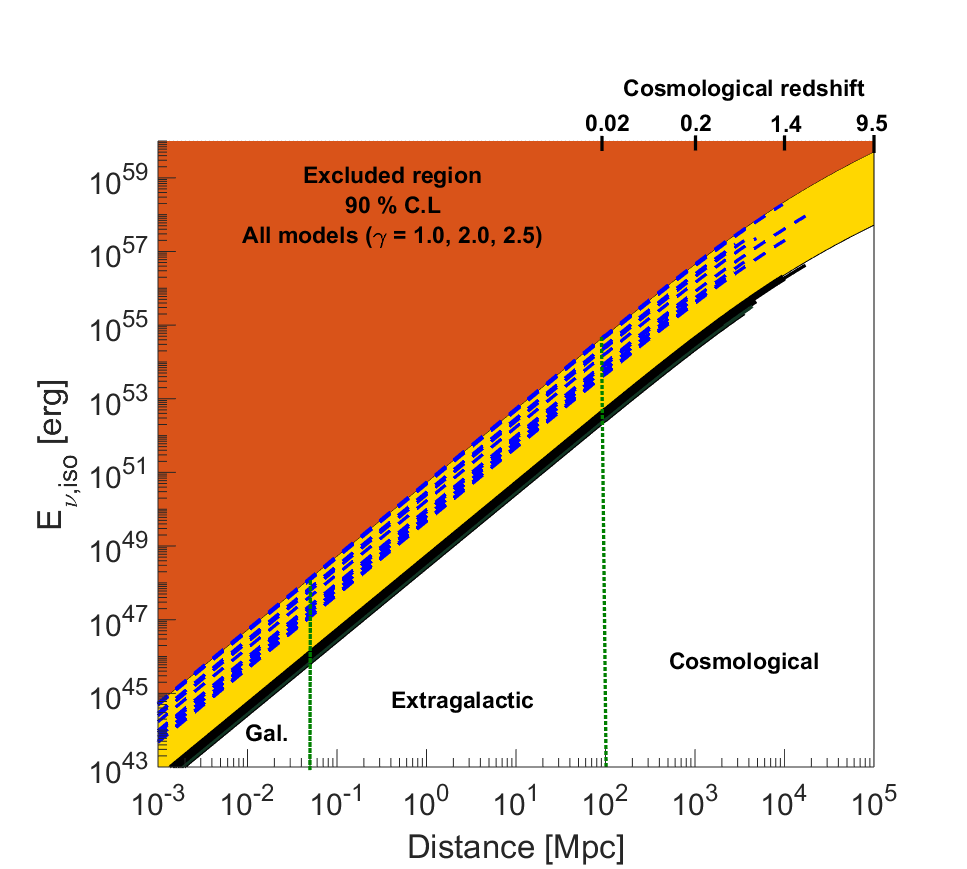}
	\caption{The 90\% confidence level upper limits on the neutrino energy released by the FRB sources. The per burst limits, assuming different neutrino power law spectra, are shown with the dashed blue lines ($\gamma$ = 1.0) and the black solid lines ($\gamma$ = 2.5). These limits are computed in the distance range $\rm{d}\in[1kpc~;D(z_{DM})]$. The red area indicates the region that is already excluded by ANTARES at the 90\% C.L. for any considered hadronic model ($\gamma$ = 1.0, 2.0, 2.5). The yellow area is only excluded by the soft spectral models ($\gamma$ = 2.0, 2.5). The white area, divided in 3 distance scenarios, is still allowed according to the ANTARES sensitivity.}
	\label{fig:Eiso_limit}
\end{figure}

The isotropic neutrino energy, $E_{\nu,iso}$, possibly released during an FRB event is an important physical property of the bursting source. It may give information on the baryonic load within the ejected outflow as well as the efficiency of the hadronic processes at work in the acceleration site nearby the progenitor source. It can be expressed as: 

\begin{equation}
E_{\nu,iso} = \frac{4\pi D(z)^2}{1+z} \cdot F_\nu
\end{equation}
where $z$ is the redshift of the source and $D(z)$ is the distance traveled by the neutrinos depending on the assumed cosmological model: 

\begin{equation}
D(z)=\frac{c}{H_0}\int_{0}^{z} \frac{(1+z')dz'}{\sqrt{\Omega_m(1+z'^3)+\Omega_\Lambda}}
\end{equation}
where $c$ is the velocity of light in vacuum and the cosmological parameters are $H_0=67.8~\rm{km~ s^{-1}~Mpc^{-1}}$, $\Omega_m = 0.308$ and $\Omega_\Lambda = 1-\Omega_m$ \citep{Planck16}. For distances in the range $\rm{d}\in[1kpc~;D(z_{DM})]$, the 90\% C.L. upper limits on $E_{\nu,iso}$ have been computed and the results are shown in figure \ref{fig:Eiso_limit} for each FRB. The excluded region in the $E_{\nu,iso}$-$D(z)$ plane for the hardest considered spectrum ($\gamma=1.0$) and the softest spectrum ($\gamma = 2.5$) are also indicated. The constraints on $E_{\nu,iso}$ obtained with the power law spectrum $\gamma$ = 2.0 are similar to that obtained with $\gamma$ = 2.5 as the two corresponding $F_{\nu}^{90\%}$ are similar.

Since the distance of these FRBs are unknown, three distance scenarios can be assumed: i) the sources are galactic or very close to our Galaxy, typically up to the distance to the Magellanic Clouds $\rm{d\le 50~kpc}$, ii) extragalactic but non-cosmological $\rm{d\in[50~kpc~;100~Mpc]}$ or iii) cosmological $\rm{d\ge 100~Mpc}$. For these three ranges of distance, the upper limits on the neutrino fluence, see figure \ref{fig:Eiso_limit}, for a $E_\nu^{-2.5}$ model, can be converted in the source rest frame by $E_{\nu,iso}^{90\%}\le [10^{43}~;10^{46}],~[10^{46}~;10^{52}]~\rm{and}~[10^{52}~;10^{57}]$ erg, respectively.

\section{Discussion}
\label{sec:Discussion}

The upper limits on the neutrino energy released by both the individual FRB sources and the whole population must be compared to the expectations of some FRB hadronic models.
\subsection{Short GRB progenitor}
\begin{figure}
	\centering
	\includegraphics[trim= 0 0 0 30,clip=true, width =0.8 \columnwidth]{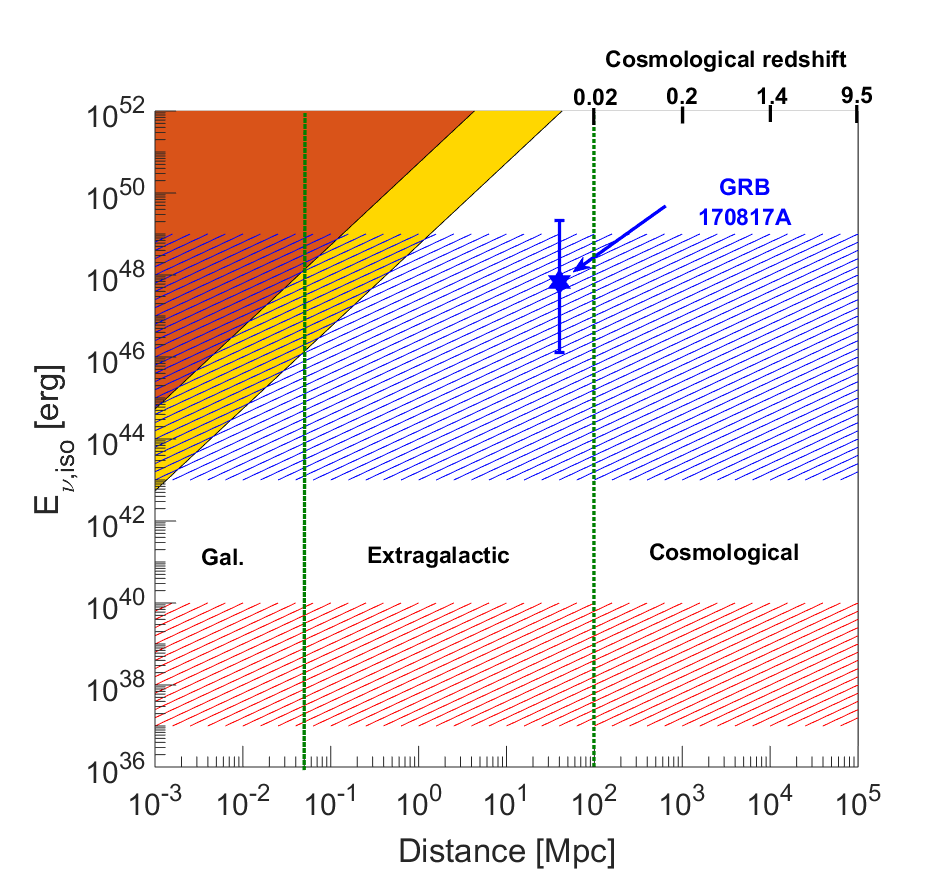}
	\caption{The $E_{\nu,iso}\rm{-distance}$ plane with the region already excluded by ANTARES for different neutrino models (red: $\gamma\ge$ 1.0, yellow : $\gamma \ge$ 2.5). The neutrino predictions from short GRBs (standard internal shock model) are represented by the blue hatched region while the magnetar/young neutron star neutrino flare expectations are shown with the red hatched area. The neutrino expectations for the short GRB GRB 170817A given by \citep{Biehl18} are also shown. The errors are due to the possible range for $f_p \rm{\in [1~;1000]}$}
	\label{fig:Eiso_limit_model}
\end{figure}

In the merger scenario, collapsing neutron stars may power an FRB at the merger time and then produce a short $\gamma$-ray burst few seconds to hundreds of seconds after \citep{Zhang14}. In the standard framework of GRBs, particles are accelerated by internal shocks within the relativistic jetted outflow and photo-hadronic processes may give rise to a burst of high-energy neutrinos \citep{Waxman97,Guetta04,Murase06,Zhang13}. The neutrino flux can be roughly scaled to the $\gamma$-ray flux and to the baryonic load in the outflow according to \citep{Zhang13} : 

\begin{equation}
E_{\nu,iso} \approx \frac{f_p}{8} \cdot (1-(1-<\chi_{p/\gamma}>)^{\tau_{p\gamma}}) \cdot E_{\gamma,iso}
\end{equation}
with $f_p$ the baryonic loading factor assumed to be preferentially in the range $f_p \in \rm{ [1~;100]}$, $\langle \chi_{p/\gamma}\rangle \sim20\%$ the fraction of the proton energy transferred to the pions, and $\tau_{p\gamma}$ the optical depth for photohadronic interactions \citep{Albert17}. For short GRBs, the isotropic $\gamma$-ray energy released is usually in the range $E_{\gamma,iso}\in [10^{47}~;10^{50}]$ erg. The short GRB 170817A associated to the binary neutron star merger event of august 2017 \citep{LIGO17} was subluminous with $E_{\gamma,iso}= 3.1\pm0.7\cdot 10^{46}$ erg integrated over an observed duration $T = (2\pm0.5)~s$ \citep{Abbott17a}. Typically, for a so-called standard short GRB\footnote{with the following parameters : the Lorentz factor $\Gamma = 300$, the gamma-ray energy $E_{\gamma,iso}\approx \rm{ 10^{50}}$ erg, the minimum variability time scale of the gamma-ray emission $t_{var} \rm{=0.01}$ s, the radius at which the p$\gamma$ interactions occur $R_{p\gamma} \approx \rm{10^{13}}$ cm and the redshift z = 0.5.}, the optical depth is $\sim 5\cdot 10^{-2}$. Based on these rough estimates, the neutrino expectations are in the range $10^{-3}\cdot E_{\gamma,iso} \lesssim E_{\nu,iso} \lesssim 0.1\cdot E_{\gamma,iso}$. As shown in figure \ref{fig:Eiso_limit_model}, the derived limits on the neutrino flux can rule out short GRB models in a very nearby environment ($\rm{d<1~kpc}$) assuming that the neutrinos are produced with an unbroken power law spectrum. Our limits can not constrain any model associating FRBs to short GRBs if the astrophysical sources are located at distances d $>$ 100 Mpc. Recent advanced hadronic models imply a broken power law spectrum for the  neutrino emission in short GRB events. Also the predictions from those models are weakly constrained by our exclusion regions. For instance, \cite{Biehl18} have computed the expected neutrino spectrum from the short GRB GRB170817A using the NeuCosmA model \citep{Hummer2010,Hummer2012} for different configurations of the jetted outflow. Considering the low luminosity jet scenario ($\Gamma = 30$, $f_p \rm{ \in [1~;1000]}$), see figure 2 given by \cite{Biehl18}, the corresponding neutrino fluences integrated over 100 TeV-100 PeV are $F_\nu \in [4.3\cdot10^{-5}~;0.07]~\rm{GeV\cdot cm^{-2}}$. At a distance of 40 Mpc and a redshift z = 0.008, this translates into $E_{\nu,iso} \in [10^{46}~;10^{49}]$ erg which is still below the ANTARES sensitivity as shown in figure \ref{fig:Eiso_limit_model}.

\subsection{Magnetar giant flare / Soft Gamma Repeater (SGR) }
In these two scenarios, the FRB event is produced by a sudden release of energy in the magnetosphere of the magnetar either driven by magnetic instabilities or high rotational loss (spin-down power). Protons may be accelerated into the polar cap regions and interact with the x-ray photon field emitted in the neutron star environment to produce high-energy neutrinos and secondary particles \citep{Zhang03}. In the first scenario, the extremely strong magnetic field ($B>10^{15}$ G) is the source of the x-ray photon field and of the particle acceleration. It corresponds to the giant flares from magnetars or the SGR models. The second scenario is related to some highly magnetised ($B\sim10^{14}$ G) neutron stars which are born with a millisecond timescale period of rotation making them able to power the particle acceleration and the subsequent high-energy neutrino emission \citep{Dey16}. In both neutron star scenarios, a very high magnetic field is required with at least $B>10^{14}$ G. For a magnetar, the typical values for the stellar radius and the magnetic field used here are $B=10^{15}$ G, $R=10$ km. The rotational period, P, can vary from hundreds of milliseconds for a very young neutron star to few seconds for slow rotating magnetars (with P $>2$ s). Based on these magnetar properties, the models of \cite{Zhang03,Dey16} predict a high-energy neutrino luminosity in the range $\rm{L_{\nu,quiescent}\in [10^{32}~;10^{35}]~erg~ s^{-1}}$ when the magnetar is in the quiescent state. For a giant flare like the one observed from SGR 1806-20 \citep{Palmer05}, the luminosity of the  x-ray/$\gamma$-ray background (with $\rm{E_\gamma = 20-30}$ keV \cite{Zhang00}) can increase by at least a factor $10^6$ in less than a second compared to the quiescent periods of the magnetar \citep{Thompson00}. 
This kind of bursting event may also produce an FRB. By scaling the typical neutrino luminosity expected from quiescent magnetars to the SGR bursting events (with typical duration for the main spike $\rm{t_{spike}}\sim$ 0.1 s \citep{Thompson00,Palmer05}), one can obtain a rough estimation of the total energy released in neutrinos during giant flares from magnetars $E_{\nu,iso}\le \rm{10^6 \cdot L_{\nu,quiescent}\cdot t_{spike} \le [10^{37};10^{40}]~erg}$. 

These estimates are also compared, in figure \ref{fig:Eiso_limit_model}, to the ANTARES neutrino upper limits. Magnetar/SGR sources are very likely weak sources of high-energy neutrino according to the models depicted above. Hence the magnetar flare origin of FRBs can not be significantly constrained on a per burst basis with the neutrino analysis presented here.\\

\subsection{Core collapse supernova environment}

Core collapse supernovae are known to produce a compact object such as a neutron star or a black hole surrounded by the material ejected from the progenitor star during the explosion, the so-called supernova remnant. \cite{Murase06,Falcke14,Ravi14, Li14} mention the possibility that cosmic-rays and high energy neutrinos may be produced by the interaction of an energetic outflow ejected by the newly born compact object with the surrounding pulsar nebula or supernova remnant at a distance R$\rm{\sim10^{15-16}~cm}$. A FRB could be also produced during this interaction or directly inside the ejected outflow. The resulting neutrino flux may be low since at such distance from the progenitor the density of the target medium for the photo-hadronic interaction is quite small. In addition, the delay between the production of the radio and the neutrino signal is not clear yet.\\

For all the hadronic models listed in this discussion, it seems that detecting a neutrino signal from single FRB sources may be difficult as most of the FRB hadronic model predictions remain orders of magnitude below the ANTARES neutrino detection threshold. However, the expected large number of FRBs over the entire sky may contribute to a diffuse flux that can be tested by a large scale neutrino telescope. This possibility is discussed in the following section.

\subsection{Contribution to the neutrino diffuse flux}

In 2013, the IceCube Collaboration reported the significant detection of a cosmic diffuse neutrino flux \citep{Aartsen13}. So far, the sources responsible for this cosmic isotropic signal have not been clearly identified. Even if the first compelling evidence of a cosmic high-energy neutrino signal from the blazar TXS 0506+056 has been claimed by \cite{IC18a,IC18b}, it is still not clear how important is the contribution of the blazars to the neutrino diffuse flux. On the contrary, the GRB contribution to this diffuse flux has already been constrained to be less than 1\% \citep{Aartsen14}. The population of fast radio bursts may also contribute significantly since their expected rate is high, $\rm{R_{FRB}\sim 10^3~day^{-1}}$ \citep{Champion16}. This hypothesis is tested here by computing the 90\% C.L. upper limit on the diffuse neutrino flux associated with FRBs. As before, the neutrino flux associated with FRB sources is assumed with a power law energy distribution with spectral indexes $\gamma$= 1.0, 2.0, 2.5. The derived diffuse upper limits depend on the assumed neutrino spectrum. Hence, 

\begin{equation}
E_\nu^2\Phi_\nu^{90\%} = \frac{1}{4\pi}\cdot \phi_{FRB}^{90\%} \cdot\frac{R_{FRB}}{N_{FRB}}~~~\rm{GeV\cdot cm^{-2} \cdot s^{-1} \cdot sr^{-1}}
\end{equation}
where $N_{FRB}$ are the number of FRBs considered in this analysis and $\phi_{FRB}^{90\%}$ is the characteristic neutrino fluence normalised to 100 TeV as defined in equation \ref{eq2b} of the combined neutrino spectrum from the 12 FRBs. The neutrino fluence limit has been computed at the 90\% confidence level by estimating the average ANTARES effective area over the 12 FRB events for the different spectral models :

\begin{equation}
\phi_{FRB}^{90\%} = \frac{2.3\cdot E_0^{-\gamma+2}}{\int \langle A_{eff}(E_\nu) \rangle \cdot E_\nu^{-\gamma}\cdot dE_\nu } ~~~\rm{GeV\cdot cm^{-2}}
\label{eq:diffuse_flux}
\end{equation}
According to the ANTARES upper limit on the individual neutrino fluxes from the 12 selected FRBs (see table \ref{tab:fluence_res}) and assuming the last updated estimate on the all-sky FRB rate $\rm{R_{FRB}\sim 1.7\cdot 10^3~day^{-1}}$ \citep{Bhandari17}, one can obtain the upper limits on the quasi diffuse flux normalised to $E_0=100$ TeV, $E_\nu^2\phi_0^{90\% }<$ 0.9, 2.0 and 0.7$\rm{\cdot 10^{-4}~GeV\cdot cm^{-2}\cdot s^{-1}\cdot sr^{-1}}$ for $E^{-1.0}$, $E^{-2.0}$ and $E^{-2.5}$ neutrino spectra respectively. 
\\
The neutrino diffuse flux observed by the IceCube Collaboration for $E_\nu>60$ TeV is at the level of $E_\nu^2\Phi_0 \sim 10^{-8}~\rm{GeV\cdot cm^{-2}\cdot s^{-1}\cdot sr^{-1}}$ normalised to $E_0=100~\rm{TeV}$ and with $\gamma = 2.46$ \citep{Aartsen15a}. In the present analysis, the derived upper limit on the diffuse flux for FRBs with $\gamma = 2.5$ is above the signal measured by IceCube by a factor $\sim$ 7300. This result is in agreement with the possibility that FRBs may originate from a wide variety of astrophysical progenitors with few of them leading to hadronic processes in their environment. In addition, according to the FRB rate mentioned above and the fact that on average one cosmic neutrino with $E_\nu>60~\rm{TeV}$ is detected every 20 days by IceCube \citep{Aartsen15a}, it appears that finally less than 1 over $\sim$ 20000 FRB could be a detectable neutrino emitter. Nevertheless, according to the IceCube analysis, the number of cosmic neutrinos at lower energy is one or two orders of magnitude larger than those with $E_\nu>60~\rm{TeV}$, depending on the spectral index of the cosmic signal and the low-energy cut-off. These cosmic neutrinos at energies below 60 TeV are hidden in the IceCube data set in the much larger sample of atmospheric neutrinos. The possibility to observe a temporal and spatial coincidence allows for a significant suppression of this background. In this paper, a selection method is presented and guarantees the 3$\sigma$ significance based on the observation of one coincident event.
If the IceCube cosmic neutrino diffuse flux is totally produced by the mechanism that induces FRBs, accounting for the neutrinos below 60 TeV, the number of neutrinos per FRB can increase by to two orders of magnitude.
This means that searches for neutrinos with IceCube or ANTARES in coincidence with hundreds of FRBs could significantly constrain such a scenario. Alternatively, the non detection of a neutrino signal from FRBs could be also due to non-hadronic production mechanisms in the FRB environment, or to the presence of a beamed jet of neutrinos. \\

Up to now, very few FRBs have been detected which strongly limits the capability of large neutrino detectors to constrain the contribution of FRBs to the neutrino diffuse flux. In the near future, many radio facilities e.g. UTMOST \citep{Caleb16,Caleb17,Bailes17}, SKA/ASKAP \citep{Johnston08,Bannister17}, CHIME \citep{Bandura14,Newburgh14,CHIME17}, Lofar \citep{vanLeeuwen14,Maan17} will increase the statistics of FRB detection up to a few per year to several hundreds per year. In the meantine, bright and very close events may be also detected (by CHIME and ASKAP, for instance) which will also increase the discovery capabilities of the large scale neutrino detectors for individual point sources.

\subsection{The complementarity of the ANTARES and IceCube detectors for FRB searches}
\begin{figure}
	\centering
	\includegraphics[trim= 0 160 0 180,clip=true,width =0.8 \columnwidth]{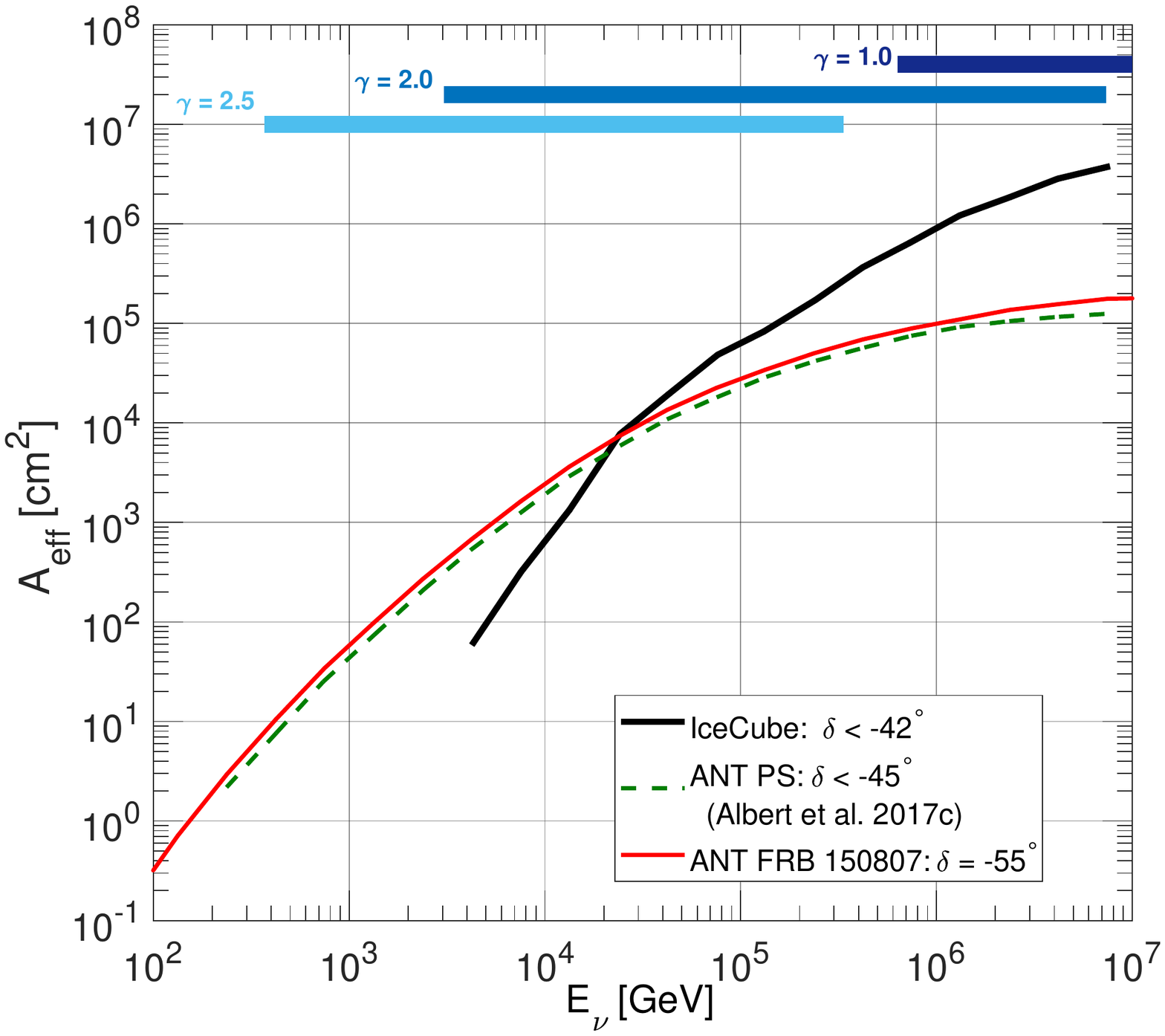}
	\caption{The ANTARES and the IceCube effective areas ($A_{eff}$) as function of the neutrino energy. The dashed dark green line is the ANTARES $A_{eff}$ computed as for the standard point source neutrino searches in ANTARES data at a declination $\delta<-45^\circ$ \citep{Albert17c}. The red line 
		illustrates the gain of about 30\% in the $A_{eff}$ ($\delta=-55^\circ$) typically achievable when searching for a transient event, such as the FRBs presented in this paper. The IceCube $A_{eff}$ computed in the last FRB analysis \citep{IC17} in the same range of Southern declinations is represented with the black line.}
	\label{fig:Aeff_all}
\end{figure}

A similar search for a high-energy neutrinos from FRBs has been performed by the IceCube Collaboration \citep{IC17}. Despite the larger detection volume with respect to the ANTARES telescope, no significant signal was found. The IceCube neutrino telescope is mostly sensitive to FRBs occurring in the Northern hemisphere where the derived upper limits on the neutrino fluence for a $E^{-2}$ spectrum are about a factor 20 more stringent than those determined by ANTARES at its maximum sensitivity (obtained for sources located in the Southern sky). However, the IceCube effective area is largely reduced for declination $\delta<-20^\circ$ with respect to positive declinations. 
Therefore, in the Southern sky, where, up to now, most of the FRBs are detected, the ANTARES telescope is still competitive with IceCube to constrain models that assume a soft spectral index such as $\gamma = 2.5$. In the Southern sky, the strongest upper limit on the neutrino fluence given by ANTARES for an FRB is a factor $\sim 1.3$ better than the one given by IceCube \citep[see the figure 7]{IC17}. Indeed, the effective area of ANTARES described in this analysis is larger than that of IceCube below $E_\nu \lesssim25~\rm{TeV}$ for the 2/3 of the Southern sky and in the large portion of the energy range where 90\% of the neutrino signal is expected for $\gamma = 2.5$. In figure \ref{fig:Aeff_all}, the ANTARES $A_{eff}$ computed at a declination $\delta= -55^\circ$ for FRB 150807 is compared to the IceCube $A_{eff}$ computed for FRB searches in the declination range $\delta \in [-90^\circ;-42^\circ]$ \citep{IC17} and illustrates the complementary between IceCube and ANTARES in terms of sky and energy coverages. Thus, in the Southern hemisphere, using both the ANTARES and the IceCube neutrino telescopes to search for transient events with soft spectra, as the one observed for the neutrino cosmic diffuse flux, maximises the discovery potential.\\
Despite the good performances of IceCube for hard neutrino spectra (high energy part of spectrum) in the Southern sky, see figure \ref{fig:Aeff_all}, a larger detector than ANTARES located in the Northern hemisphere is required to improve the sensitivity of the neutrino telescopes to sources located in the Southern hemisphere. The next generation of the large-scale\footnote{with a detection volume of the order of the km$^3$} high-energy neutrino detector, KM3NeT/ARCA \citep{KM3NET_LOI16}, is now under construction in the Mediteranean sea and will be fully operational in the upcoming years. With KM3NeT/ARCA, an unprecedented sensitivity to the high-energy neutrino flux should be obtained at Southern declinations. In the next few years, combined analysis between the KM3NeT/ARCA and the IceCube detectors will provide the most sensitive and homogeneous coverage of the neutrino sky ever reached for energies $E_\nu >1~\rm{TeV}$.

\section{Conclusions}
\label{Conclusion}
Fast radio bursts are candidate sources of efficient particle acceleration as they may release a great amount of energy in a short timescale, similar to the short gamma-ray bursts. The question whether a hadronic component is injected in the energetic outflow has been investigated in this paper by directly searching for high energy neutrinos in coincidence with 12 FRBs using ANTARES data in the period 2013-2017. No significant coincident neutrino signal was found. The 90\% confidence level upper limits on the neutrino fluence has been derived per burst and for the whole population as well as the neutrino energy released. These limits are not stringent enough to significantly constrain the prediction of some FRB hadronic models, e.g. merger events, magnetar flares, especially if these sources are located at cosmological distances. FRBs could be weak sources of high energy neutrinos, but because of their high rate in the Universe, the signal from the whole population may be detectable as a diffuse neutrino flux. So far, the lack of FRB statistics does not allow to firmly test this hypothesis since the detection of at least a hundred of FRBs are required to bring significant constraints. The upcoming first observations of KM3NeT/ARCA, the next generation of large-scale high-energy neutrino telescope in the Northern hemisphere, will also permit to strongly improve the constraints on the fluence per burst and the FRB contribution to the cosmic neutrino diffuse flux. Recently, new facilities such as UTMOST, CHIME, SKA/ASKAP, have emerged with high discovery capabilities of tens of FRBs per month (against $\rm{\sim3-4}$ per year as during the last ten years) covering not only the Southern sky. This will allow to fully exploit the available and future neutrino detector capabilities for FRBs detected in the North sky with the IceCube and IceCube-Gen2 as well as ANTARES and KM3NeT/ARCA at Southern declinations (note also that the KM3NeT/ARCA dectector should have at least comparable performances than the current IceCube detector in the Northern sky for declinations $\delta\lesssim +40^\circ$). In addition, these radio facilities may be able to observe bright FRBs at close distance ($\rm{d<100~Mpc}$), which will enhance the probabilty of a multimessenger detection at high energies for an individual FRB. Finally, more accurate models describing the neutrino production associated with FRBs will greatly help to refine the constraints on the neutrino fluence and energy released.


\section*{Acknowledgements}
The authors acknowledge the financial support of the funding agencies:
Centre National de la Recherche Scientifique (CNRS), Commissariat \`a
l'\'ener\-gie atomique et aux \'energies alternatives (CEA),
Commission Europ\'eenne (FEDER fund and Marie Curie Program),
Institut Universitaire de France (IUF), IdEx program and UnivEarthS
Labex program at Sorbonne Paris Cit\'e (ANR-10-LABX-0023 and
ANR-11-IDEX-0005-02), Labex OCEVU (ANR-11-LABX-0060) and the
A*MIDEX project (ANR-11-IDEX-0001-02),
R\'egion \^Ile-de-France (DIM-ACAV), R\'egion
Alsace (contrat CPER), R\'egion Provence-Alpes-C\^ote d'Azur,
D\'e\-par\-tement du Var and Ville de La
Seyne-sur-Mer, France;
Bundesministerium f\"ur Bildung und Forschung
(BMBF), Germany; 
Istituto Nazionale di Fisica Nucleare (INFN), Italy;
Stichting voor Fundamenteel Onderzoek der Materie (FOM), Nederlandse
organisatie voor Wetenschappelijk Onderzoek (NWO), the Netherlands;
Council of the President of the Russian Federation for young
scientists and leading scientific schools supporting grants, Russia;
National Authority for Scientific Research (ANCS), Romania;
Mi\-nis\-te\-rio de Econom\'{\i}a y Competitividad (MINECO):
Plan Estatal de Investigaci\'{o}n (refs. FPA2015-65150-C3-1-P, -2-P and -3-P, (MINECO/FEDER)), Severo Ochoa Centre of Excellence and MultiDark Consolider (MINECO), and Prometeo and Grisol\'{i}a programs (Generalitat
Valenciana), Spain; 
Ministry of Higher Education, Scientific Research and Professional Training, Morocco.
We also acknowledge the technical support of Ifremer, AIM and Foselev Marine
for the sea operation and the CC-IN2P3 for the computing facilities.


\end{document}